# Substitutional Doped GeSe: Tunable Oxidative States with Strain Engineering

Zheng Shu,[a] Yongqing Cai[b*]

[a]Center for Data Science, Institute of Collaborative Innovation, University of Macau, Taipa, Macau, China

[b]Joint Key Laboratory of the Ministry of Education, Institute of Applied Physics and Materials Engineering, University of Macau, Taipa, Macau, China

Email: yongqingcai@um.edu.mo

**Abstract**

Layered chalcogenide materials have a wealth of nanoelectronics applications like resistive switching and energy-harvesting such as photocatalyst owing to rich electronic, orbital, and lattice excitations. In this work, we explore monochalcogenide germanium selenide GeSe with respect to substitutional doping with 13 metallic cations by using first-principles calculations. Typical dopants including s-shell (alkali elements Li and Na), p-shell (Al, Pb and Bi), 3d (Fe, Cu, Co and Ni), 4d (Pd and Ag) and 5d (Au and Pt) elements are systematically examined. Amongst all the cationic dopants, Al with the highest oxidation states, implying a high mobility driven by electric field, and Al-doped GeSe may be a promising candidate for novel resistive switching devices. We show that there exist many localized induced states in the band gap of GeSe upon doping Fe, Co, or Ni, while for Cu, Ag, and Au cases there is no such states in the gap. The Ag and Cu are + 0.27 and + 0.35 charged respectively and the positive charges are beneficial for field-driven motion in GeSe. In contrast, Au is slightly negatively charged renders Au-doped GeSe a promising photocatalyst and enhanced surface plasmon. Moreover, we explore the coexistence of dopant and strain in GeSe and find dynamical adjustments of localized states in GeSe with levels successive shifting upward/downward with strain. This induces dynamic oxidative states of the dopants under strain which should be quite popular in composites where motion of metal adatoms causes significant deformation.

**Introduction**

The discovery of graphene and other two-dimensional (2D) layered crystal materials, such as black phosphorus (BP), stanene, transition metal dichalcogenides (TMDS), etc, has triggered great interests in various phenomena in atomically thin space[1-8]. These 2D materials bring about novel electronic and photonic performance different from their counterpart bulk materials due to their electrons, holes and phonons are confined in a small limited space[9-11]. As a result of unique physical properties, 2D materials are promising for some innovative applications such as advanced catalyst and nanoelectronic devices[12,13]. Recently, layered group-IV monochalcogenides (GeS, GeSe, SnS and SnSe), a 2D family with isostructural and isoelectronic analogue of phosphorene have attracted great attentions. Due to their high stability and intriguing electronic properties, such chalcogenides compounds are potential candidates for resistive switching devices[14-16] and electrochemical memory cells[17,18].

Germanium selenide (GeSe), which has highly in-plane anisotropic electronic and optical properties due to its unique "puckered" symmetry structure, has aroused extensively attention[19] for optoelectronics owing to its environmentally more friendly and biocompatible than other chalcogenides. Bulk GeSe has an intrinsically p-type conducting characteristics (similar to phosphorene) with a closely placed direct-indirect band gaps around 1.1 eV[20]. GeSe films have an excellent absorption under near infrared (NIR) excitation[21,22], an optimal gap for a single junction solar cells [23-25], efficient phototransistor[26,27], and a highly sensitive to NIR light irradiation for photodetector[28-30]. 2D GeSe monolayer was successfully fabricated onto $Si/SiO_2$ substrates by using laser thinning, vapor transport and deposition techniques[30,31]. Anisotropic conductance inherent in GeSe is ideal platform for developing high-efficiency plasmonic devices where the resonance frequency can be continuously tuned by light polarization direction[32,33]. GeSe monolayer has also been proven to have giant piezoelectricity[34] and thermoelectrics[35] allowing energy harvesting. Twisted bilayer GeSe was reported to show intriguing Moiré patterns with the crossover from two to one dimension excitations[36]. More recently, GeSe has been found to be an ideal platform for laser[37] and novel nanoelectronic devices such as memristors[38,39]. Few-layer GeSe sheets, together with proper ionic particles like Ag, are used as the ionic and electronic conductor, where high and low resistance with ratio up to $10^4$ have been achieved for information storage[38]. An Ag/GeSe/TiN memristor for brain-inspired neuromorphic applications was fabricated and an electronic synapse was realized through controlling the electromigration and diffusion of Ag atoms within the GeSe flakes[40]. Therefore, understanding the behaviors metallic particles within the GeSe is highly important for such applications. The dopants can create significant distortion and strains[41] in the lattice while such effects are still unknown. Understanding the oxidation states of the dopants under external stimuli is ultra-important. However, related information is still missing and unknown.

In this work, we investigate the coupling of the doping and strain effects on the electronic properties of GeSe by using first-principles calculations on the basis of density functional theory (DFT). We are particularly interested in the oxidative states of such dopants which are highly correlated with the chemical and novel electronic applications (i.e. resistive switching). We investigate the localized levels and charged states of these dopants which are found to vary with lattice strain. This suggests a dynamic oxidative states of the dopants under strain which can provide useful hints in memory devices and energy applications.

**Computational method and details**

The structure optimization and electronic properties are performed using the Vienna ab initio simulation package (VASP)[42], which is based on the framework of spin-polarized density functional theory. We adopt a supercell (5 × 4 × 1) of monolayer GeSe structure. The cutoff energy of 400 eV and *k* points of 3×3 grid are used. Furthermore, a vacuum layer with thickness of 15 Å is adopted to avoid spurious interaction of images under the periodic boundary condition. The structures are fully relaxed until the forces on atoms are smaller than 0.01 eV/Å. We also calculate the binding energy $E_b$ to evaluate the stability of X-GeSe system. The equation for $E_b$ of doping system is defined as:

$$E_b = E_{X+\text{GeSe}} - E_X - E_{\text{vacancy}}$$

where *X* represents doped element, $E_{X+\text{GeSe}}$ denotes total energy of *X* doped GeSe monolayer, $E_X$ denotes energy of single atom *X*, $E_{\text{vacancy}}$ denotes the total energy of GeSe monolayer with one Ge atomic vacancy.

**Results and discussion**

In this work, typical cationic dopants, with following typical electronic configurations: s-shell (alkali elements Li and Na), p-shell (Al, Pb and Bi), 3d (Fe, Cu, Co and Ni), 4d (Pd and Ag) and 5d (Au and Pt), are systematically studied by substituting one of the germanium atom (forming $X_{Ge}$ type of defect where $X$ is the dopant) in the supercell. The doping concentration is ~3%. The new bond length, formation energy and magnetic properties are calculated and listed in Table 1. Most doped species form stable hybridized bonds to atoms of GeSe except Li, Na dopants.

**Table 1.** Binding energy ($E_b$ in eV), magnetic moment ($M$ in μB), charge transfer $\Delta q$ from doping cations to GeSe calculated by Bader charge analysis, and the $X$-Se, $X$-Ge bond length ($B_{X\text{-Se}}$, $B_{X\text{-Ge}}$) where $X$ represents doped elements for substitutional doping ($X_{Ge}$) in GeSe. We only list the range of bond length when two or more bonds are formed. Note a positive $\Delta q$ of the dopants indicates a net positive charge formed in the dopant with a transfer of electrons to the host sheet of GeSe.

| Dopant (X) | $E_b$ (eV) | $M$ (μB) | $\Delta q$ (e) | $B_{X–Se}$ (Å) | $B_{X–Ge}$ (Å) |
|---|---|---|---|---|---|
| Li | -1.16 | 0.29 | + 0.99 | 2.42 | 2.80-2.98 |
| Na | -3.27 | 0 | + 0.99 | 2.88-2.94 | |
| Ag | -2.97 | 0 | + 0.27 | 2.58-2.64 | 3.20 |

| Au | -3.03 | 0 | - 0.06 | 2.51-2.67 | |
| Cu | -4.36 | 0 | + 0.35 | 2.36-2.40 | |
| Pd | -4.13 | 0 | + 0.04 | 2.43-2.51 | 2.57 |
| Pt | -5.66 | 0 | - 0.24 | 2.42-2.53 | 2.69-2.73 |
| Fe | -5.09 | 4.00 | + 0.84 | 2.37-2.41 | |
| Co | -5.93 | 2.96 | + 0.65 | 2.34-2.38 | |
| Ni | -5.19 | 1.91 | + 0.44 | 2.31-2.34 | |
| Bi | -2.83 | 0.998 | + 0.87 | 2.80 | |
| Pb | -4.51 | 0 | + 1.88 | 2.83-3.34 | |
| Al | -4.64 | 0.90 | + 3.00 | 2.45-2.55 | |

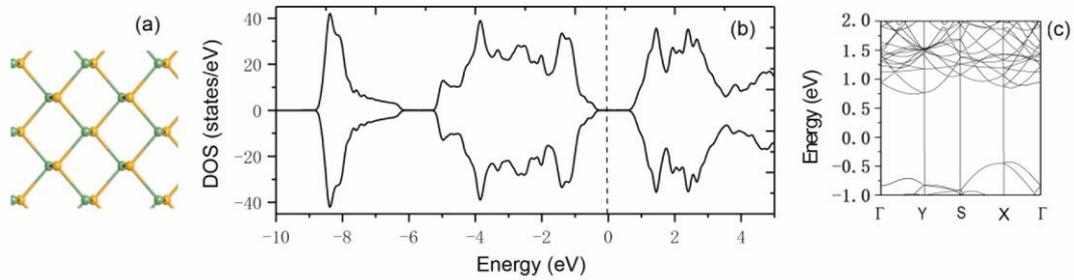

**Fig. 1** The atomic structure (a), DOS (b), band structure (c) of pristine GeSe monolayer of a 5 × 4 supercell. Note the zigzag (armchair) direction is along the y (x) direction.

**Electronic properties of pristine GeSe:** GeSe sheet adopts a puckered structure similar to phosphorene with forming Ge-Se zigzag chains aligned along the armchair direction (Fig. 1a). Our optimized lattice constants of unit cell of monolayer GeSe are a= 3.58 Å, b= 3.91 Å, consistent with previous studies[43,44]. The atomic configuration, density of states (DOS) and band structure of pristine GeSe (for a 5 × 4 supercell) are plotted in Fig. 1. Pristine GeSe monolayer exhibits a non-magnetic semiconducting characteristic with no spin splitting (Fig. 1b).

It is shown that GeSe monolayer has an indirect band gap (1.17 eV), which is close to previous reported values (1.16 eV and 1.22 eV) in Ref. 45, 46 respectively. However, these two papers demonstrate that GeSe monolayer has a direct band gap. But a recent research[47] shows that monolayer GeSe has an indirect band gap (1.13 eV) which is consistent with our work. By the way, Ref. 48 shows there is only a very small energy distinction between the direct and indirect gap. As shown in Fig. 1c, in addition to the conduction band minimum (CBM) at Y point, there is a sub-valley at X point with around 0.1 eV slightly above the CBM, directly above valence band maximum (VBM). This suggests that GeSe is likely to transit from indirect to direct gap under moderate external disturbance like strain and electric fields which may lead to the cross of positions of CBM and sub-valley. A proper band engineering with taking advantage of this subtle transition will lead to significant enhancement of the photon absorption.

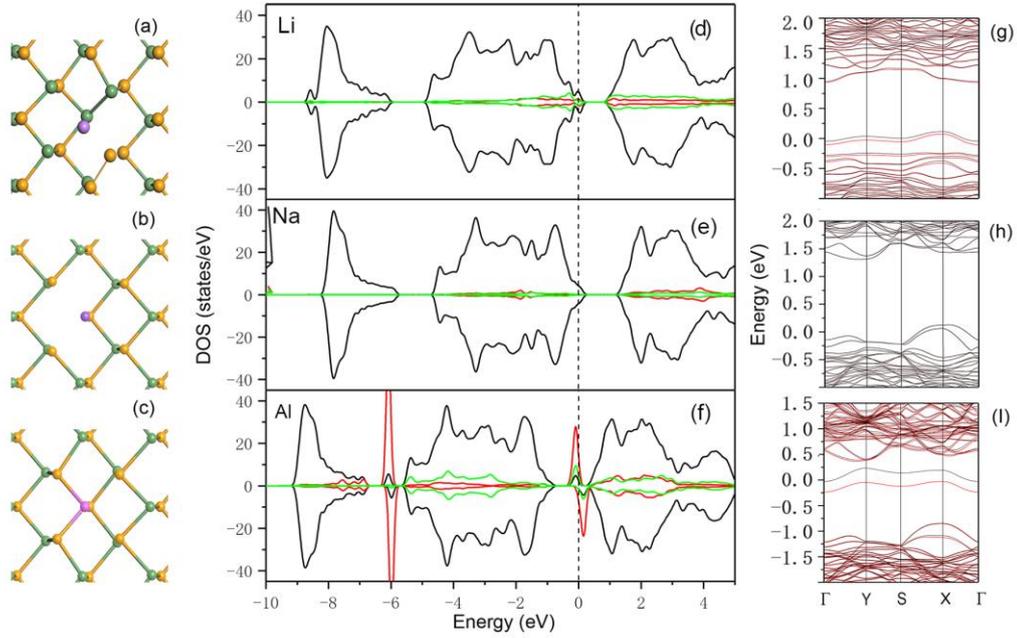

**Fig. 2** The bonding configurations (a)-(c) for Li, Na and Al doped GeSe monolayer, respective, and DOS (d)-(f). The black, red, green and blue curves represent the total DOS and the local density of states (LDOS) of the s-, p- and d-orbitals of doped atom, respectively. Note that the values of LDOS of the Li, Na (Al) dopants are enlarged by a scale of 100 (10) for better comparison. The band structures (g)-(i) correspond to Li, Na and Al doped GeSe monolayer, respectively. For band structures, the black (red) curves represent spin-down (spin-up) states.

*Li, Na and Al doped GeSe monolayer:* We give an attention to doped monolayer GeSe with Li, Na and Al atoms that have no electrons in their d-orbitals (Fig. 2). Alkali species doped 2D materials have attracted a great deal of interests for electrochemical applications (i.e. hydrogen storage or lithium battery) because of its weak binding of alkali species on surface and high surface-volume ratio[49-51]. Table 1 compiles the energetics of Li and Na doping process in monolayer GeSe with the $E_b$ of -1.16 and -3.27 eV, respectively. For alkali group elements Li and Na atoms, during structural optimization the dopants are found to be highly mobile in the sheet by indication of large displacement of their positions from the initial structure. This could be attributed to its unique electronic structure with 1s shell. Our Bader charge analysis shows that this outer 1s electron of Li is almost completely transferred to the GeSe sheet. This creates fully ionic Li, Na-X (X=Se, Ge) bonds (Fig. 2a and b) which are highly flexible and dynamic.

It should be noted that although Li and Na almost donate all of its outer 1s electron to the GeSe, the Fermi level ($E_F$) of the hybrid Li/Na-GeSe system is still below the VBM of GeSe, indicating a p-type conduction for such type of substitutional doping. The reason is that the substitutional doping ($Li_{Ge}$ or $Na_{Ge}$) is essentially a composite defect ($Li_i+V_{Ge}$ or $Na_i+V_{Ge}$) consisting of interstitial Li or Na ($Li_i$, $Na_i$) and Ge vacancy ($V_{Ge}$). The creation of a $V_{Ge}$ leaves two holes states and the Li or Na donates single electron which only partially compensates it, thus there still has one hole in each $Li_{Ge}$ or $Na_{Ge}$. As shown in Fig. 2g, the donating electron from Li at the $V_{Ge}$ causes slight splitting of the bands and pushes up of the $V_{Ge}$ related bands (crossing $E_F$), while no such effect are found in the Na case which should be attributed to the more localized electrons in Li than in Na.

In the Al case, the Al atom forms Al-Se bonds with three adjacent selenium atoms and the band length ranges from 2.45-2.55 Å (Fig. 2c). The $E_b$ is -4.64 eV indicating a much stronger adsorption than Li and Na. Bader charge analysis shows that three electrons are transferred from Al to GeSe and Al is +3 charged. The transferred three electrons allow the fully occupation of the two holes left from the $V_{Ge}$, and the other excess electron transferred to the GeSe host and pushes the $E_F$ above the VBM, as shown by the DOS and band structure for Al doped GeSe monolayer in Fig. 2f and i, respectively. This demonstrates that $Al_{Ge}$ dopants induces n-type carriers. In addition, Al dopant creates two spin-split mid-gap states around the $E_F$ with one is empty and the other occupied.

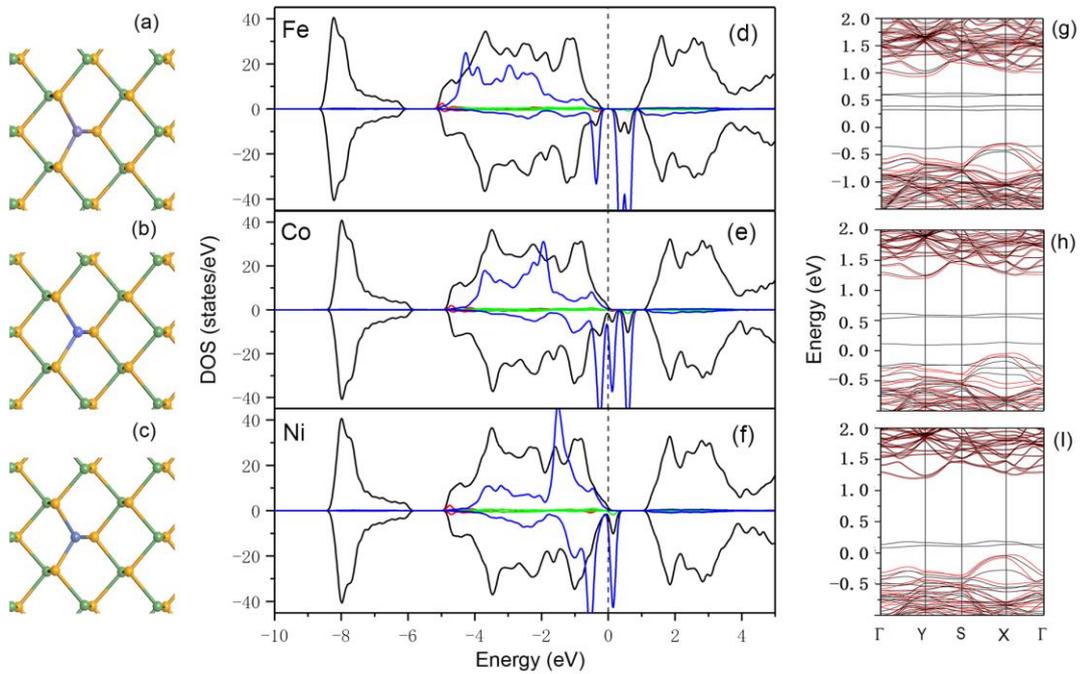

**Fig. 3** The bonding configurations (a)-(c) for Fe, Co and Al doped GeSe monolayer, respective, and DOS (d)-(f). The black, red, green, blue curves

represent the total DOS and the local density of states (LDOS) of the s-, p- and d-orbitals of doped atom, respectively. The values of LDOS of these dopants are enlarged by a scale of 10. The band structures (g)-(i) correspond to Fe, Co and Ni doped GeSe monolayer, respectively.

***Fe, Co and Ni doped GeSe monolayer:*** We next consider the Fe, Co, and Ni which are typical magnetic elements. Our calculation shows that these three VIIIB-group elements have a strong affinity with three adjacent selenium atoms, and binding energy $E_b$ with monolayer GeSe is -5.09, -5.93 and -5.19 eV, respectively. This is similar to their strong interaction with the cousin oxygen elements. All the bonding number of these three dopants is 3, and these three doped monolayer GeSe have the strongest $E_b$ compared with other dopants considered in this work (Fig. 3a-c).

As shown in the DOS plot (Fig. 3d-f), there exist significant localized dopant induced levels in the band gap of GeSe. For Fe dopant, there are five doping levels appearing in the gap of GeSe and the magnetic moment of Fe doped system is 4.00 μB, which is the maximum value among the other 13 dopants. Four doping levels of them are above the $E_F$ and the occupied residual level touches the VBM. For Co dopant, there are three doping levels appearing and the magnetic moment of Co doped system is 2.96 μB. For Ni dopant, there are two doping levels appearing and the magnetic moment is 1.91 μB. The reason that these three dopants have relatively large magnetic moment is exchange splitting in their 3d orbitals.

In addition, as shown in the band structure (Fig. 3g-i), the successive decreasing number of the defective levels follows 4 (Fe) > 3 (Co) > 2 (Ni), and there is an overall downward trend relative to the VBM from Fe, Co to Ni. This could be rooted in the different electronegativity with Fe < Co < Ni. The larger value of the electronegativity, the deeper the electronic levels from the vacuum energy. For Fe with the smallest electronegativity, it has the best ability of donating its electrons to the GeSe. Indeed, as shown in Table 1, the amount of transferred electrons follows Fe (+ 0.84) > Co (+ 0.65) > Ni (+0.44).

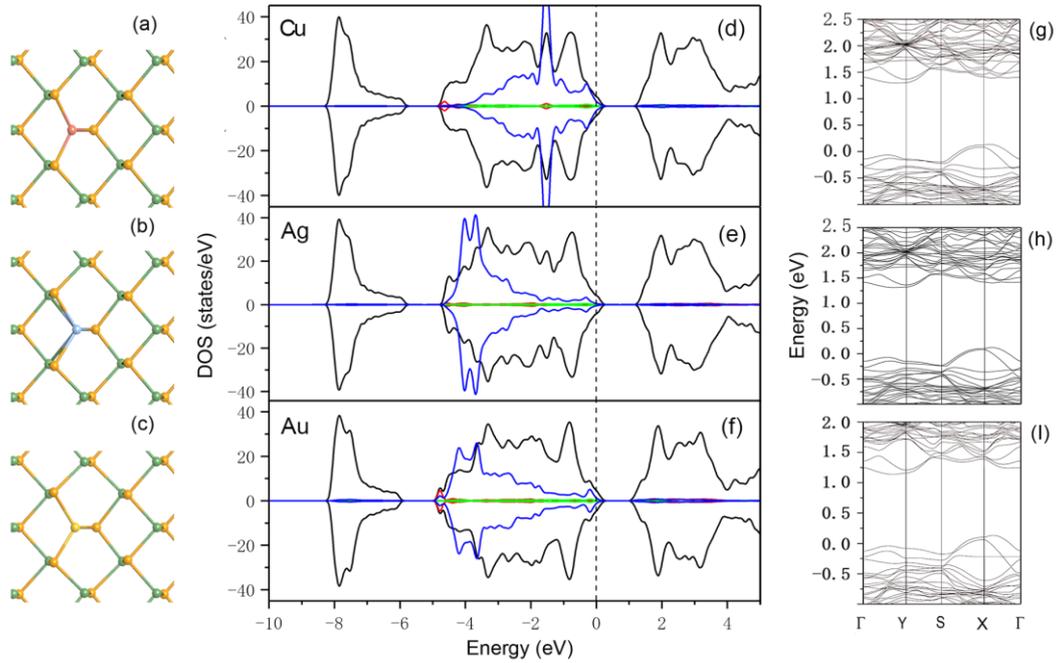

**Fig. 4** The bonding configurations (a)-(c) for Cu, Ag and Au doped GeSe

monolayer, respective, and DOS (d)-(f). The black, red, green, blue curves represent the total DOS and the local density of states (LDOS) of the s-, p- and d-orbitals of doped atom, respectively. Note that the values of LDOS of these three dopants are enlarged by a scale of 10. The band structures (g)-(i) correspond to Cu, Ag and Au doped GeSe monolayer, respectively.

***Cu, Ag and Au doped GeSe monolayer:*** The Cu, Ag and Au atoms have $nd^{10}(n+1)s^1$ electronic configurations. Their $n$d states are fully occupied and $(n + 1)$s state is singly occupied. The bonding configurations of these dopants are listed in Fig. 4a-c. The binding energy $E_b$ with monolayer GeSe is -4.36, -2.97 and -3.03 eV for Cu, Ag, and Au, respectively. It can be seen that the structures of Cu and Au doped monolayer GeSe are analogous to original monolayer GeSe and only undertake a minor relaxation. Both Cu and Au doped atoms form X-Se bonds with three adjacent selenium atoms. The bond lengths of Cu-Se bonds and Au-Se bonds are 2.36-2.40 Å and 2.51-2.70 Å, respectively. However, for Ag case, it not only forms three Ag-Se bonds but also two Ag-Ge bonds with two adjacent germanium atoms. The bond lengths of Ag-Se bonds and Ag-Ge bonds are 3.58-2.64 Å and 3.20 Å, respectively. The respective electronic DOS are plotted in Fig. 4d-f. As we can see, there exist significant resonance between the d states with states of GeSe in the bottom (top) part of the valence band for Ag and Au (Cu). Thus the main mechanism for their interaction with GeSe monolayer is through d-orbital hybridization.

The band structures of the Cu, Ag, and Au doped GeSe are shown in Fig.4g-i. Compared with the pristine GeSe, the Cu, Ag, and Au do not induce any additional states in the band gap. Moreover, there is minor distortion of the bands of the GeSe upon the substitutional doping of such elements which is quite similar to the Na-doped GeSe. Similarly, the $E_F$ is slightly below the VBM indicative of a p-type conduction. For these three metal doped systems, no magnetic moments (Table 1) and spin splittings are observed. Bader charge analysis shows that Ag and Cu are + 0.27 and + 0.35 charged, respectively. According to previous studies[52,53] on Ag, Cu and Au doped oxide, such adatoms are charged due to the transfer of outermost singly occupied s state to the hosting materials. Interestingly, in the case of Au doped GeSe, the Au is slightly negatively charged. This implies that in addition to the depletion of the 1s electron of Au, there exist another mechanism of backdonation of electrons to Au which is highly likely to be related to the hybridization of Au-d state with GeSe. Considering the premier performance of the GeSe under near infrared (NIR) excitation and absorption[21], the Au-doped GeSe may be a promising photocatalyst and enhanced surface plasmon [54] due to the good charge transfer to Au and the benign states associated with doped Au. Our results of showing the positive charge of doped Ag and Cu could also help to the explanation of the dissolution of such metallic particles in the active GeSe[16].

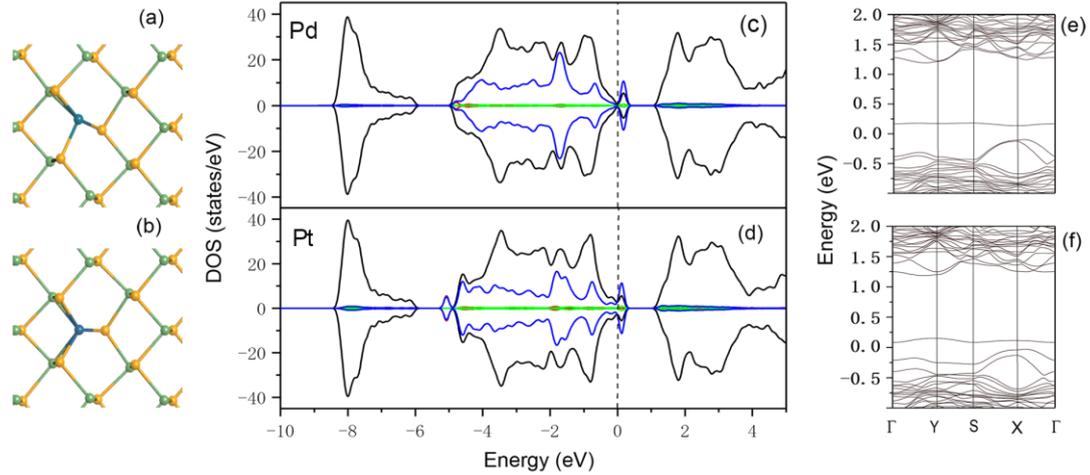

**Fig. 5** The bonding configurations (a, b) of Pd and Pt doped GeSe monolayer, respectively, and DOS (c, d). The black, red, green, blue curves represent the total DOS and the local density of states (LDOS) of the s-, p- and d-orbitals of doped atom. The values of LDOS of two dopants are enlarged by a scale of 10. The band structures (e, f) correspond to Pd and Pt doped GeSe, respectively.

*Pd and Pt doped GeSe monolayer*: Next, we explore noble metal (Pd and Pt) doped GeSe monolayer. The bonding configurations for these two dopants are shown in Fig. 5a and b, respectively. Surprisingly, the substitutions of Pd and Pt trigger relative complicated structures and deformation of GeSe. For the Pd case, it forms three Pd-Se bonds with three adjacent Se atoms and one Pd-Ge bond with an adjacent germanium atom. The bond lengths of Pd-Se bonds and Pd-Ge bonds are 2.43-2.52 Å and 2.57 Å, respectively. For the Pt case, it forms three Pt-Se bonds with three adjacent Se atoms and two Pd-Ge bonds with two adjacent Ge atoms like Ag dopant. The bond lengths of Pt-Se bonds and Pt-Ge bonds are 2.40-2.53 Å and 2.69-2.73 Å, respectively. These bonding geometries are consistent with a relatively strong $E_b$ of Pd (-4.13 eV) and Pt (-5.66 eV) as listed in Table 1. In particular, the strong adsorption of Pt atom with GeSe implies in real device with Pt electrode, there must exist strong electrode-GeSe interaction and some of the Pt atoms may diffuse into GeSe, reducing the contacting resistance and barrier. The effect would be more significant when there are remarkable Ge vacancies in the GeSe before the deposition of Pt electrode.

Figure 5 c-f present the DOS and band structure for Pd and Pt doped GeSe monolayer, respectively. For both Pd and Pt systems, no magnetic moments is found. Both Pd and Pt dopants create an empty in-gap state located around 0.3 and 0.2 eV higher above the VBM, respectively. Thus novel optical properties may appear for Pd and Pt doped GeSe monolayer due to formation of this doping level. Notably, the introduction of Pd causes negligible distortion of the bands of GeSe. In contrast, the Pt induces significant changes of the band dispersions of GeSe, in particularly for the bands around VBM. This could be explained by the much stronger interaction in the Pt than Pd according to our above energetics analysis.

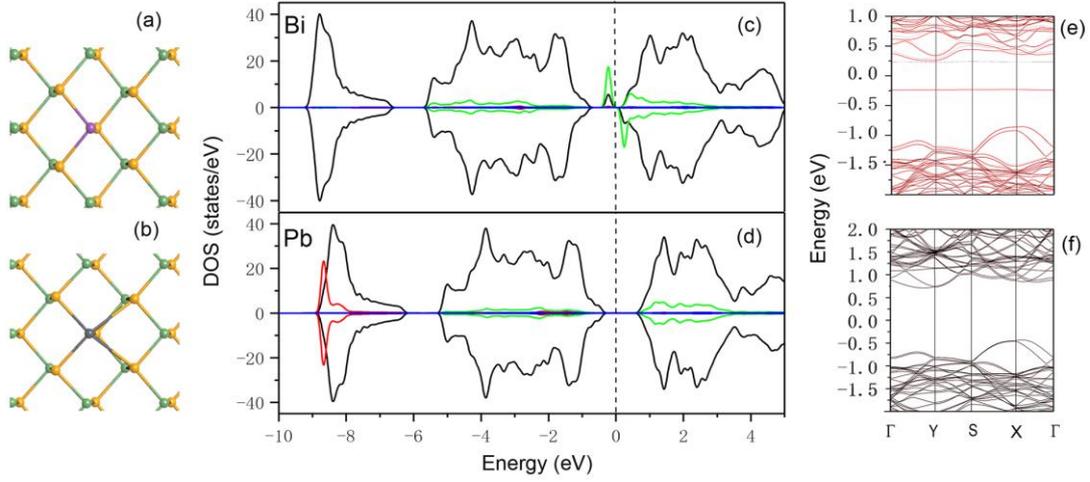

**Fig. 6** The bonding configurations (a, b) of Bi and Pb doped GeSe monolayer, respectively, and DOS (c, d). The black, red, green, blue curves represent the total DOS and the local density of states (LDOS) of the s-, p- and d-orbitals of doped atom. The values of LDOS of two dopants are enlarged by a scale of 10. The band structures (e, f) correspond to Bi and Pb doped GeSe, respectively.

*Bi and Pb doped GeSe monolayer*: Finally, we give a discussion to Bi and Pb doped monolayer GeSe. Their bonding configurations are plotted in the Fig. 6a and b, respectively. The Bi atom forms Bi-Se bonds with two adjacent Se atoms while in the case of Pb there forms Pb-Se bonds with five adjacent Se atoms. For Bi, the bond length of two Bi-Se bonds is 2.80 Å. The binding energy $E_b$ of the two doped elements is listed in Table 1, the calculated results are -2.83 and -4.51 eV for Bi and Pb, respectively. The reason of this bonding geometry is that Bi atom has lone pair electrons and thus a moderate $E_b$ of Bi just lower than Li (-1.16 eV). The bond length of five Pb-Se bonds ranges from 2.83-3.34 Å. Figure 6 c-f show the DOS and band structure for Bi and Pb doped GeSe monolayer, respectively. For Bi, notably, significant hybridizations of Bi-p state and GeSe are found in the bottom part of the conduction band. There are two localized levels within the band gap of Pb-doped GeSe: one empty shallow level touching the CBM and one occupied deep level around 0.5 eV below CBM with opposite spin polarization. This also leads to 1.00 μB magnetic moment due to the Bi dopant. In contrast, for Pb doped system no magnetic moment is observed and there is no dopant-induced state in the band gap of GeSe monolayer. The band structure of Pb-doped GeSe has negligible changes compared with pristine GeSe, showing robustness against the Pb dopant.

### *3.7 Strain-induced variation of band structure*

Manipulating electronic and magnetic properties of two-dimensional (2D) materials with effective approach is vital to explore the applications of 2D materials in many fields. Strain effects are proved to be an effective approach of tuning electronic and other properties[55-59]. In addition, examining the deformation of the lattice on the electronic properties is critical as 2D materials are highly likely to be subjected to external mechanical deformation. In this section, we investigate the variation of band structure and magnetic moment for pristine GeSe monolayer and substitutional doped GeSe monolayer using first-principles calculation.

The motivation of studying the strain effect on the doped GeSe is that presence of strain is unavoidable in real applications like resistive switching devices based on chalcogenides[38-40] where strain is created during the insertion and depletion of these conducting ions. Here, only pristine GeSe, and typical metallic dopants like Au, Ag, Cu, Pd and Pt are chosen for strain analysis of doped GeSe. We apply moderate biaxial compressive and tensile strains ranging from -2%-8% and examine the variation of the band structures. We are particularly interested in the changes of those dopant induced states with strain. Indeed, under a series of different strains, some intriguing characteristics of pristine and doped GeSe monolayer triggered by strain are found.

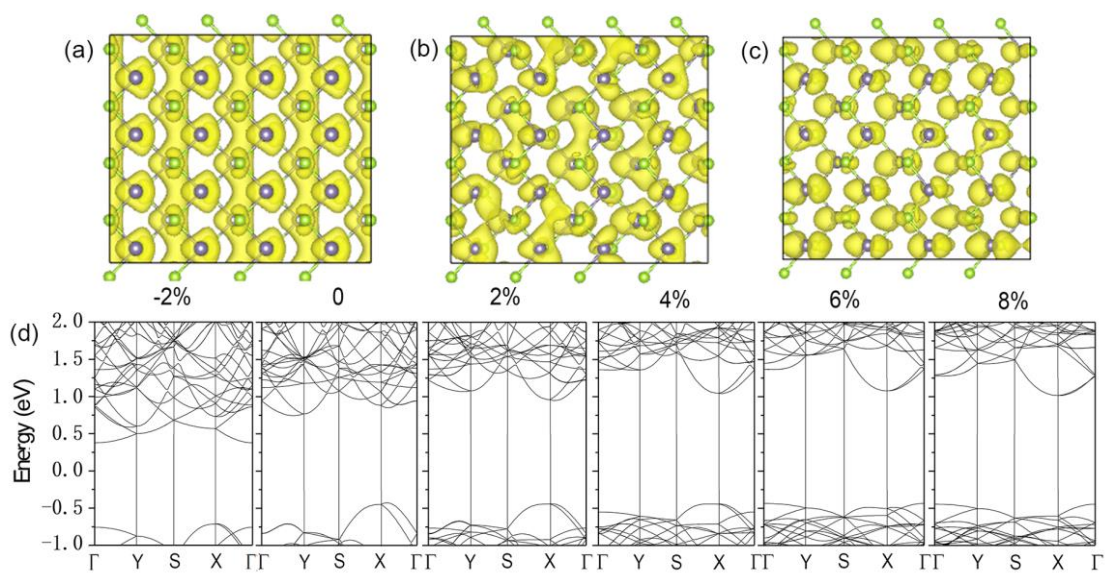

**Fig. 7** (a) Electronic densities of CBM of pristine GeSe monolayer under 0, 4% and 8% bi-axial strains. (d) Evolution of band structures of pristine GeSe monolayer under -2%, 0, 2%, 4%, 6% and 8% bi-axial strains.

***Strain engineering of pristine GeSe monolayer***: Firstly, we investigate the strain engineering of pristine GeSe monolayer. As shown in Figure 7a-c, the states of CBM become more localized with increasing tensile strains. Compared with zero strain, there are trends of dimerization of CBM states at 4%. We only consider moderate compressive strain of -2% since larger strains will lead to rippling and wrinkles of the 2D material. For applying -2% strain, the GeSe still maintains an indirect band structure with the band gap reduces to 1.09 eV from 1.17 eV of zero strain. Compared with zero strain case, both CBM and VBM drop significantly (~0.7 eV) toward low-energy direction with the CBM shifting from the Y point to the Γ point. Although the VBM still locates at the X point, the energy difference to the second maximum point (at Γ) becomes tiny to the scale of tens of meV. Therefore, direct light excitation via state transitions at Γ should be more prominent under moderate compressive strain.

For tensile strains, the band gap changes to 1.39, 1.48, 1.51 and 1.46 eV for 2%, 4%, 6%, and 8%, respectively. As shown in Fig. 7d, the bands become flattened with tensile strains due to the gradual localization of the electrons. For the valence bands, the distance from the VBM relative to $E_f$ almost has no significant changes while the maximum point shifts from X point to the Γ point starting from 6% and multiple band crossings occur. In contrast, the CBM shifts upwardly from zero to 4% strain and levels off afterwards. The position of CBM evolves from Y to X point from 2% onward. Therefore, from zero to 2% strain, the band structure occurs an indirect to direct transformation, and evolves back to indirect gap again at ~ 6% strain. Our results are in good agreement with previous studies[55-59].

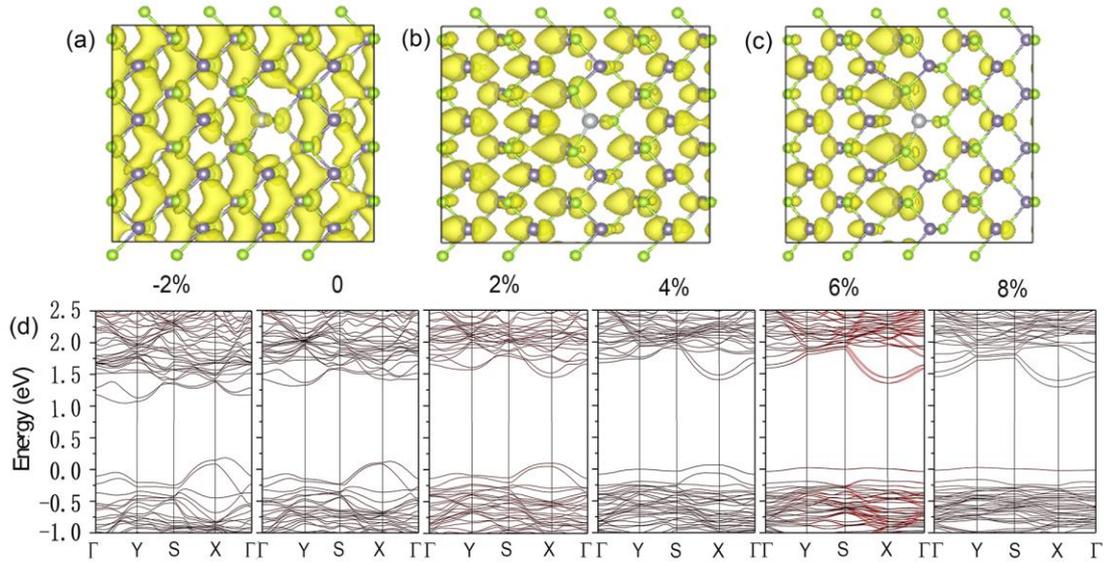

**Fig. 8** (a) Electronic densities of CBM of Ag-doped GeSe monolayer under 0, 4% and 8% bi-axial strains. (d) Evolution of band structures of Ag-doped GeSe monolayer under -2%, 0, 2%, 4%, 6% and 8% bi-axial strains. Black (red) lines represent spin-up (down) states.

***Strain engineering of Ag doped GeSe monolayer:*** Next, strain effects for Ag doped GeSe monolayer are systematically studied. Ag/GeSe has recently proven to be an important resistive switching devices, where significant strains are induced during the insertion of Ag atoms in the GeSe[38,40]. As shown in Figure 8a-c, for the CBM state, less density is distributed around the Ag dopant compared with the remaining part of GeSe under zero strain. Notably, with increasing strain there are accumulated charges around the vicinity of Ag dopant. Such enhanced states are antibonding which may lead to photo-induced diffusion, growth, and coalescence of Ag particles in GeSe under light excitation with appropriate wavelength. The incorporation of Ag shifts the $E_f$ below the VBM at zero strain as indicated in Fig. 8d. Under zero-strain condition, the band structure is analogous to pristine GeSe monolayer that CBM appears at Y point while VBM appear at X point. With the applied strain increases, the valence band becomes flatter. When the tensile strain is up to 4%, there is a mid-gap state appearing around the Fermi level which should be related to the intrinsic states of Ge vacancy, partially passivated by the Ag dopant. Our calculation suggests that there are dynamical adjustment of localized states with the levels shifting upward/downward upon the adding/releasing of the strain. Such strains are highly likely to be triggered by the motion of the Ag particles across GeSe sheets in resistive switching devices.

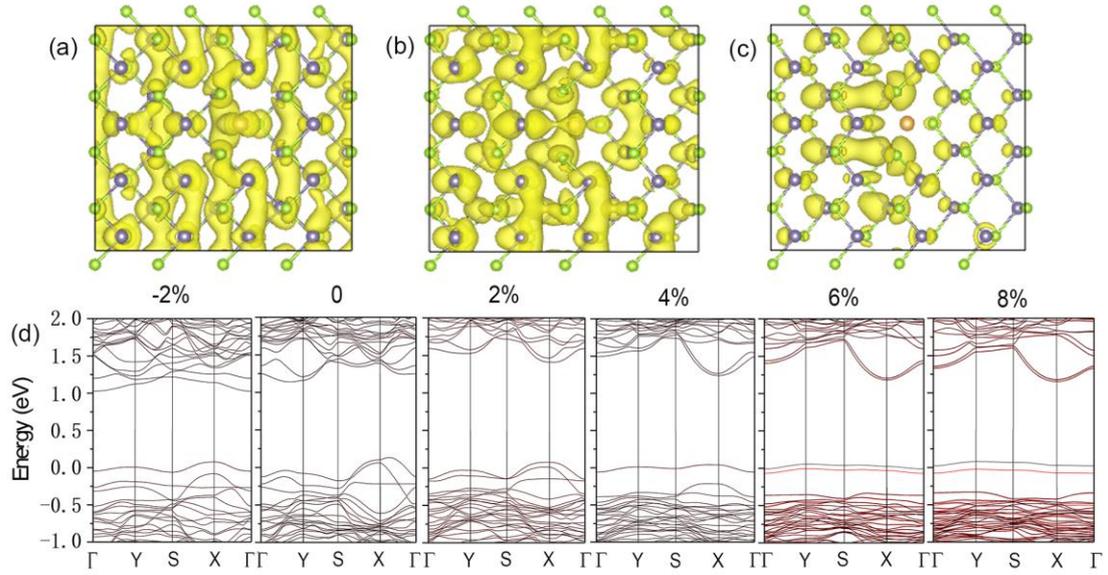

**Fig. 9** (a) Electronic densities of CBM of Au-doped GeSe monolayer under 0, 4% and 8% bi-axial strains. (d) Evolution of band structures of Au-doped GeSe monolayer under -2%, 0, 2%, 4%, 6% and 8% bi-axial strains. Black (red) lines represent spin-up (down) states.

***Strain engineering of Au doped GeSe monolayer***: For the Au-doped GeSe, the CBM contains significant component of Au with overwhelming distribution along zigzag direction. With increasing tensile strain, the pattern gradually evolves into armchair characteristics and more states are localized around Au, implying potential light-enhanced plasma excitation of Au-doped GeSe (Fig. 9a-c). Under -2% compression strain, there is a mid-gap level close to the VBM and around $E_f$ (Fig. 9d). Under tensile strain, the valence bands become flatter and an in-gap state forms when strain is up to 4%. This doping level becomes spin split when the strain exceeds 6%. This phenomenon justifies that 6% tensile strain induce magnetization and this value is 0.32 μB through calculation. With the strain up to 8%, the magnetic moment increases to 0.51 μB. The intriguing fact implies that strain engineering is an effect approach to make doped GeSe monolayer yielding magnetism. The spin-split should be results from the modulated fractional electron hopping between Au and the Se deficient GeSe. Such dynamic charge transfer implies the multiple oxidative states of Au in GeSe systems which allows accumulations of photo-excited carriers in loaded Au. In particular, as shown in Fig. 9, starting from 6%, there are three localized states in the band gap, either from the $V_{Ge}$ or the Au. These additional states well above the valence bands of GeSe form as the traps of photo-excited carriers intrinsically allows more resonant excitations.

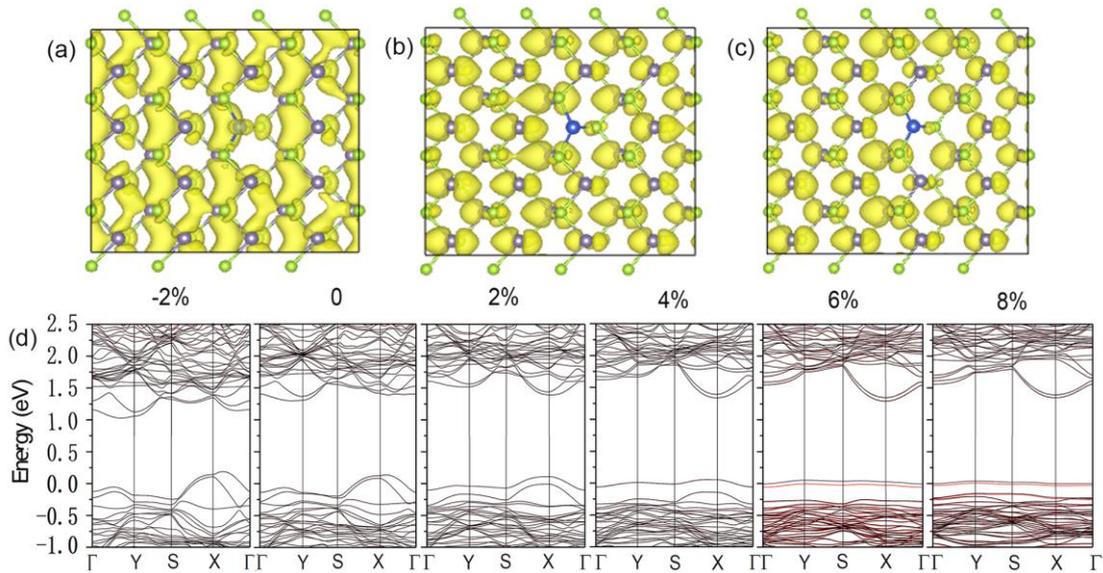

**Fig. 10** (a) Electronic densities of CBM of Cu-doped GeSe monolayer under 0, 4% and 8% bi-axial strains. (d) Evolution of band structures of Cu-doped GeSe monolayer under -2%, 0, 2%, 4%, 6% and 8% bi-axial strains. Black (red) lines represent spin-up (down) states.

***Strain engineering of Cu doped GeSe monolayer***: Similar to other cases, CBM states of Cu-doped GeSe become gradually localized with increasing tensile strain. However, different from Ag and Au cases, the CBM states of Cu-doped GeSe under high tensile deformation still largely maintain uniform. This could be due to a weaker strain field around the Cu-dopant due to a much smaller nuclei radius of Cu than Ag and Au (Fig. 10a-c). As shown in Fig. 10d, upon applying negative strains, the gap narrows in Cu doped GeSe. For tensile strains, the valence band becomes flatter with increasing tensile strains and induces a localized state, highly likely from $V_{Ge}$, when strain is up to 4% liked Au dopant. This level splits into two states, one is spin-up while the other is spin-down when the strain is up to 6%. This phenomenon justifies that 6% tensile strain induces magnetic moment of 0.33 µB. The overall behaviors of Cu doped GeSe is quite similar to the Au case when strains are applied albeit the localized levels in the Cu case under high strains being closer to the VBM. This difference could be due to a much smaller nuclei of Cu which relieves the outward expansion of the surrounding Se atoms compared with Au atom.

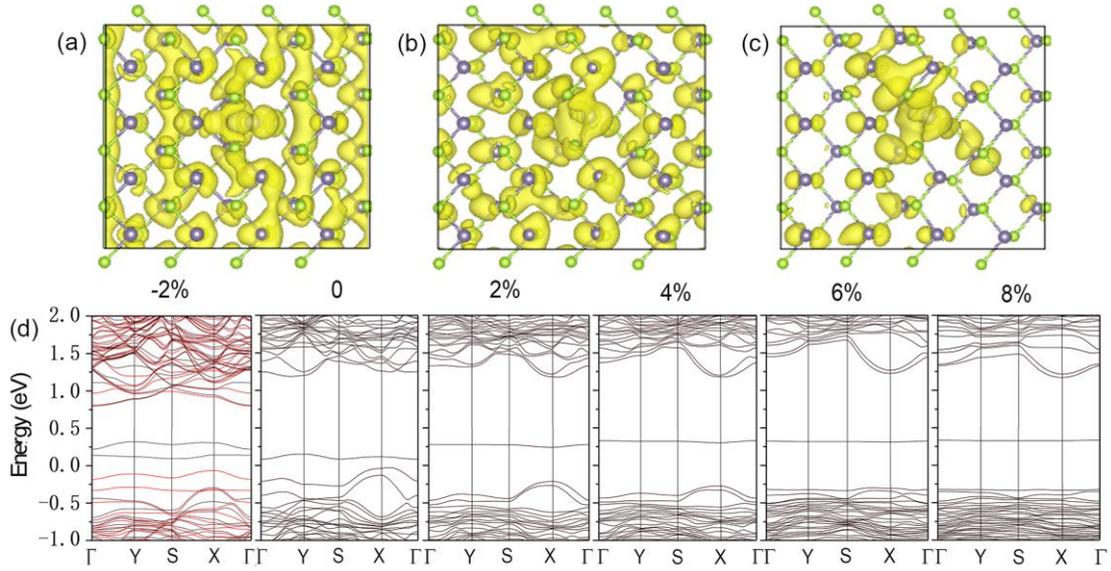

**Fig. 11** (a) Electronic densities of CBM of Pt-doped GeSe monolayer under 0, 4% and 8% bi-axial strains. (d) Evolution of band structures of Pt-doped GeSe monolayer under -2%, 0, 2%, 4%, 6% and 8% bi-axial strains. Black (red) lines represent spin-up (down) states.

***Strain engineering of Pt doped GeSe monolayer***: For Pt doped case, the response with strain is quite intriguing and different from previous cases. As shown in Fig.11a-c, the CBM states of Pt-doped GeSe become predominantly localized around Pt under high strains. As shown in Fig. 11d, there is an empty level associated with the Pt dopant around 0.2 eV above VBM without strain. Different from other dopants, the in-gap levels are associated with the dopant rather than the $V_{Ge}$. When we apply -2% compression strain, this mid-gap level splits into two levels. The bands are spin polarized and two additional localized levels in the gap with opposite spin occurring underneath the $E_f$. The Pt doped GeSe monolayer yields magnetic moment (1.88 μB) under -2% compression strain. For tensile strains, Pt dopant has the same trend as previous dopants where the valence band becomes flatter with strain increasing. However, the in-gap state greatly jumps from 0.2 eV (zero strain) to 0.5 eV (2%) and remains for higher strains.

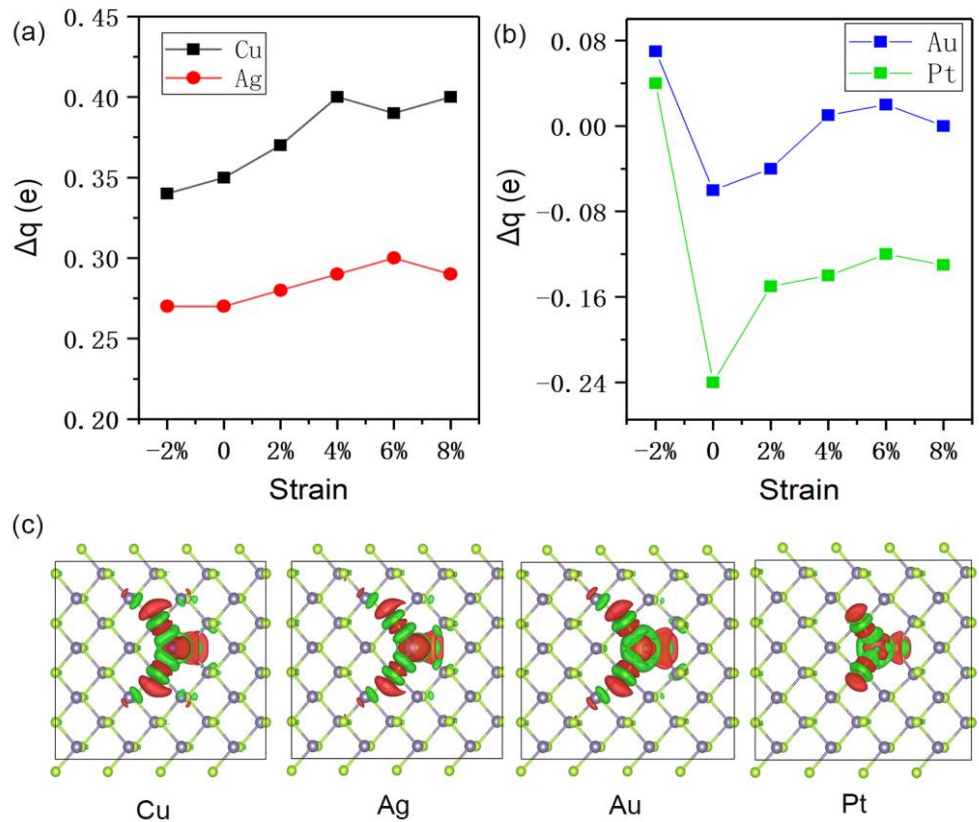

**Fig. 12** Evolution of charged states of doped atoms Ag and Cu (a), and Au and Pt (b) as a function of strain. (c) Isosurface of differential charge density for Cu, Ag, Au and Pt-doped GeSe. The red (green) color represents the electronic charge accumulation (depletion) upon doping.

**Modulated oxidative states of cationic dopants with strain:**

Finally, we perform Bader charge analysis on the charged state of each dopants. As shown in Fig. 12a and b, the Cu (around +0.35 e) and Ag (around +0.27 e) carry positive charges for all the strains, while Pt holds negative charges for all the tensile strains (range from -0.24 to -0.12 e) and slightly positive charge for -2% strain. Au is the most neutral dopant amongst all the examined elements. It is slightly negatively charged from 0 to 2% strain (around -0.06 e) and slightly positively charged for other strains. All the dopants become more oxidized with more electrons are transferred to the GeSe with more tensile strains being applied. The different amount of charge transfer of different dopants can be largely understood by the electronegativity of the metal dopants. The smaller the electronegativity the larger the amount of the electron transfer from the dopants to the GeSe. On the contrary, the elements with a larger electronegativity would have less tendency to transfer electrons to GeSe. For instance, the electronegativity of Ag is 1.93 (in Pauling scale) is larger than Cu of 1.90, which accounts for the more positively charged state of Cu (+0.28 e) than Ag (+0.38 e) in GeSe. In contrast, the Au and Pt has the biggest electronegativity of 2.54 and 2.28, respectively, while their charged states are even negatively charged under zero strain as shown in Figure 12. Due to the much larger electronegativity that Ag and Cu[52], they are less oxidized in GeSe. Nevertheless, they also possess an overall trend of increasing oxidation with strain. Differential charge density plot (Fig. 12c) for typical dopants i.e. Cu, Ag, Au and Pt shows that due to chemical hybridization significant amounts of charges from/to the dopants are transferred to/from the Se-p orbital.

For Ag and Cu which are normally adopted in resistive switch devices, the higher degree of charges carried by the dopants under larger tensile strains implies that stronger forces by electric field will be exerted on these moving particles in highly stretched films. Consider the higher degree for deformation at the frontier area of newborn filament, our result of higher charged dopants in more stretched region uncovers the acceleration mechanism of the moving particles and growth of the filament in these devices[38,40]. Our results also demonstrate that the moving particles seems not encompass a single charged state in such devices but with a series of oxidative states. The presence of the multiple oxidative states is also a key requirement for satisfying catalytic functionality of such metallic doped semiconductor. These results are consistent with the experimentally observed localized states near the Fermi level which are enhanced after metallic doping[41,60].

Finally we would like to comment on the occurrence of the strain in such resistive devices. Very large nonuniform compressive or tensile strains can be triggered by the intercalation of the nanoparticles. During the SET/RESET operation of resistive switching devices, this will create the dynamical adjustment of conducting filament. The latter normally takes place through the growth of nanoparticles like Ag or Cu in the semiconducting hosts. Rippling of the 2D materials during ionic insertion or extraction of the particles. Moreover, very large nonuniform compressive or tensile strains up to 10% can be formed in the convex and concave region of the ripples of phosphorene[61] which is isostructural of GeSe. Compressive strains could be triggered by co-factors like temperature reduction, substrate (i.e. flexible substrate), electric field induced resistive switching, etc.

## 4. Conclusion

In summary, we investigate the substitutional metal doped GeSe monolayer by using first-principles calculations based on density functional theory. The energetics and electronic properties of these doped GeSe monolayers with respect to the localized states are systematically explored in this paper. Our work indicates that there exist significant amounts of localized induced states in the band gap of Fe, Co, Ni-doped GeSe. In contrast, the Cu, Ag, and Au do not induce any additional states in the band gap and only cause minor distortion of the bands of the host. Moreover, there are dynamical adjustment of localized states in GeSe with the levels shifting upward/downward upon the adding/releasing of the strain. This induces dynamic oxidative states of the dopants under strain which is ubiquitous in resistive switches and novel nanoelectronic devices.

**Acknowledgements**

This work is supported by the University of Macau (SRG2019-00179-IAPME) and the Science and Technology Development Fund from Macau SAR (FDCT-0163/2019/A3). This work was performed in part at the High Performance Computing Cluster (HPCC) which is supported by Information and Communication Technology Office (ICTO) of the University of Macau.